\documentclass{article}
\begin{document}
\title{Anisotropic
cosmological models with non-minimally coupled magnetic field}
\author{
Alexander B. Balakin\footnote{Electronic address: Alexander.Balakin@ksu.ru}\\
Department of General Relativity and Gravitation  \\
Kazan State University, 420008 Kazan, Russia\\and\\
Winfried Zimdahl\footnote{Electronic address:
zimdahl@thp.uni-koeln.de}\\
Institut f\"ur Theoretische Physik, Universit\"at zu  K\"oln\\
D-50937 K\"oln, Germany }
\date{\today}
\maketitle

\begin{abstract}
Motivated by the structure of one-loop vacuum polarization effects
in curved spacetime we discuss a non-minimal extension of the
Einstein-Maxwell equations. This formalism is applied to Bianchi I
models with magnetic field. We obtain several exact solutions of
the non-minimal system including those which describe an
isotropization process. We show that there are inflationary
solutions in which the cosmological constant is determined by the
non-minimal coupling parameters. Furthermore, we find an isotropic
de Sitter solution characterized by a ``screening" of the magnetic
field as a consequence of the non-minimal coupling.
\end{abstract}
\vspace{0.8cm}
\vspace{0.5cm}

\section{Introduction}

The Einstein-Maxwell theory has been a subject of investigations
since long \cite{ExactSolutions}. As far as cosmological
implications are concerned, the possible role of a primordial
magnetic field has attracted particular interest. Studying the
impact of a magnetic field on the dynamical evaluation requires
anisotropic cosmological models. A general discussion of this type
of models with magnetic field and references to early activities
along this line may be found in \cite{ZelNov}.  By now there are
strong limits on the current magnitude of such a field
\cite{Barrow} which seems to render magnetic fields on
cosmological scales unimportant at the present stage of the cosmic
evolution. This does not imply, however, that a magnetic field did
not influence the dynamics of the early Universe. Moreover,
quantum effects are expected to become relevant at early cosmic
stages. Quantum electrodynamical consideration show that vacuum
polarization effects in curved spacetime give rise to non-minimal
modifications of the (minimal) Einstein-Maxwell Lagrangian
\cite{Drum}. The investigation of a non-minimal coupling of
gravity with electromagnetic fields was initiated by Prasanna
\cite{Prasa1}. Prasanna introduced the additional invariant
$R^{ikmn}F_{ik}F_{mn}$ ($R^{ikmn}$ is the Riemann tensor, $F_{ik}$
is the Maxwell tensor) into the Lagrangian for the
gravito-electromagnetic system and obtained a non-minimal
one-parameter modification of the Einstein-Maxwell equations
\cite{Prasa2} .  Novello and Salim \cite{Novello1} included the
(gauge-dependent) terms $R A^k A_k$ and $R^{ik}A_i A_k$ in the
Lagrangian ($A_k$ is the electromagnetic potential four-vector,
$R^{ik}$ is the Ricci tensor and $R$ is the curvature scalar). A
qualitatively new step has then been made by Drummond and Hathrell
\cite{Drum} by calculating quantum-electrodynamical one-loop
corrections in curved spacetime. The Lagrangian of such a theory
contains the three  U(1) gauge-invariant scalars
$R^{ikmn}F_{ik}F_{mn}$, $R^{ik}g^{mn}F_{im}F_{kn}$ and
$RF_{mn}F^{mn}$ with coefficients proportional to the square of
the Compton wavelength of the electron. Subsequently, a
non-minimal coupling of gravity and electromagnetism has been
discussed by a number of authors in different settings
\cite{Acci1,Novello2,Souza,Go,Acci2,Turner,MH,Novello3,Acci3,Mohanty,Tess}.
Non-minimally extended theories were used as a framework to
discuss potential limitations of the equivalence principle
\cite{Lafrance,Prasa3,Solanki}). A further quantum
electrodynamical motivation of the use of the generalized Maxwell
equations can be found in \cite{CollKost,Kost1,Kost2}. The effect
of birefringence induced by curvature, first discussed in
\cite{Drum}, and some of its consequences for electrodynamic
systems have been investigated for the pp-wave background in
\cite{Balakin1,Balakin2,Balakin3,Balakin4,Balakin5}. A curvature
force has been introduced to describe the accelerated expansion of
the universe \cite{Antifr,NJP}. Non-minimal interactions in which
torsion is coupled to the electromagnetic field were studied in
\cite{Hehl1,Hehl2} (see also \cite{Hehl3} for a review). Finally,
we mention a mediated non-minimal coupling in which the scalar
Higgs field $\phi$ is coupled to gravity via a $\xi \phi^2 R$ term
and to a Yang-Mills potential $ A_k$ by $\phi^2 A_k A^k$
\cite{Bij}.

Most of the investigations so far were devoted to the analysis of
the non-minimally modified Maxwell equations on a given
background. With the exceptions \cite{Acci1} and \cite{Novello3}
the impact of the non-minimal coupling on the gravitational field
has been outside the focus of interest. However, one may expect
that the rich structure of such type of theories gives rise to
novel features in the gravitational dynamics as well. Our purpose
here is to demonstrate this aspect for a specific class of
non-minimal interactions in Bianchi I cosmological models. These
interactions are modelled according to the already mentioned
general structure obtained in \cite{Drum} as the result of quantum
electrodynamical calculations. While the coupling constants are
fixed in \cite{Drum}, they will be considered as arbitrary,
constant parameters in our analysis (cf. \cite{Turner}). On this
basis we shall first obtain the general set of equations for a
non-minimally extended Einstein-Maxwell system with linear
electrodynamics (For a non-linear, non-minimal extension of the
Einstein-Maxwell theory see \cite{BaLe05}). Then we specify this
set to the homogenous but anisotropic case of Bianchi I
cosmological models with magnetic field and a matter component
with generally anisotropic equations of state. We obtain simple
exact solutions for several choices of the non-minimal coupling
parameters. Among them are solutions with axial symmetry which
isotropize in the long time limit. As a specific feature of
inflationary solutions of the non-minimal theory we find direct
relations between a cosmological constant and the non-minimal
coupling parameters. There exists an isotropic de Sitter solution
with this property as well for which the corresponding set of
parameters makes the gravitational dynamics independent of the
(non-vanishing) magnetic field.

The paper is organized as follows. In section \ref{non-minimal} we
present the general formalism of the non-minimally extended set of
gravito-electromagnetic field equations, based on \cite{Drum}. The
(linear) electrodynamical field equations are obtained an
discussed in section \ref{Electrodynamic equations}. Section
\ref{Gravitational field} is devoted to the gravitational field
equations in the general case. In section \ref{Anisotropic models}
the material of the previous sections is applied to the Bianchi I
geometry. Several particular models are then studied in section
\ref{Particular models}. Section \ref{Discussion} provides a
summary of the paper.

\section{Non-minimal coupling of gravity and electromagnetism}
\label{non-minimal}

\subsection{General formalism}

A non-minimal extension of the Einstein-Maxwell theory can be
derived from the action functional
\begin{equation}
S[g,A] = \int d^4 x \sqrt{-g} \ L \label{act}
\end{equation}
with the Lagrangian
\begin{equation}
L = \frac{R + 2 \Lambda}{\kappa} + {\cal L}_{{\rm matter}} +
\frac{1}{2} F_{mn}F^{mn} + \frac{1}{2} {\cal R}^{ikmn}
F_{ik}F_{mn} \,. \label{lag}
\end{equation}
Here, $\Lambda$ is a cosmological constant, $g$ is the determinant
of the metric tensor $g_{ik}$, the constant $\kappa$ is equal to
$\kappa = \frac{8\pi G}{c^4}$ where $G$ is Newton´s gravitational
constant. The quantity ${\cal L}_{{\rm matter}}$ is the Lagrangian
of neutral matter and $F_{ik}$ is the Maxwell tensor $F_{ik} =
\nabla_{i}A_{k} - \nabla_{k}A_{i}$, where $\nabla_{i}$ denotes the
covariant derivative and $A_{i}$ is the four potential. The last
term describes a ($U(1)$ gauge invariant) non-minimal coupling
between gravity and electromagnetism, mediated through the tensor
\\
\begin{eqnarray}
{\cal R}^{ikmn} &\equiv & \frac{q_1}{2}(g^{im}g^{kn} {-}
g^{in}g^{km}) R \nonumber\\ && {+} \frac{q_2}{2} (R^{im}g^{kn} {-}
R^{in}g^{km} {+} R^{kn}g^{im} {-} R^{km}g^{in}) {+} q_3 R^{ikmn}
\,, \label{sus}
\end{eqnarray}
\\
where $q_1$, $q_2$ and $q_3$ are phenomenological coupling
constants with the dimension $[{\rm length}]^{2}$. This structure
of the coupling is motivated by quantum electrodynamical
calculations of vacuum polarization effects in curved spacetime by
Drummond and Hathrell \cite{Drum}. While $q_1$, $q_2$ and $q_3$
have definite values in \cite{Drum}, we assume them to be
arbitrary constant parameters in our analysis. The choice
(\ref{lag}), (\ref{sus}) is a generalization of previous
non-minimal modifications of Maxwell`s theory. The case
$q_1=q_2=0$ was investigated in \cite{Prasa1,Prasa2}. For
$q_2=q_3=0$ one obtains a model considered in \cite{Novello3}.

The tensor ${\cal R}^{ikmn}$ has the same symmetry properties as
the Riemann tensor $R^{ikmn}$. Contraction yields
\\
\begin{eqnarray}
g_{kn}{\cal R}^{ikmn} &=&  R^{im}(q_2 + q_3) + \frac{1}{2}R
g^{im}(3 q_1 + q_2) \,,\nonumber\\  g_{kn}g_{im}{\cal R}^{ikmn}
&=& R (6 q_1 + 3 q_2 + q_3) \,. \label{sustrace}
\end{eqnarray}
\\
The case of a vanishing trace which will be of interest in later
applications is characterized by
\\
\begin{equation}
g_{kn}g_{im}{\cal R}^{ikmn} = 0\quad\Rightarrow\quad 6 q_1 + 3 q_2
+ q_3 = 0 \,. \label{zerotrace}
\end{equation}
\\

\section{Electrodynamic equations}
\label{Electrodynamic equations}

The equations of non-minimal electrodynamics are obtained by
varying  the action functional with the Lagrangian (\ref{lag})
with respect to the four-potential $A_i$ of the electromagnetic
field. They are of the standard form
\begin{equation}
\nabla_{k} H^{ik} = 0  \,, \quad \nabla_{k} F^{*ik} = 0  \,,
\label{eld1}
\end{equation}
where the induction tensor $H^{ik}$ is given by
\begin{equation}
H^{ik} \equiv F^{ik} + {\cal R}^{ikmn} F_{mn}  \,. \label{eld2}
\end{equation}
This relation has the structure of a constitutive law in which
${\cal R}^{ikmn}$ plays the role of a susceptibility tensor.

\subsection{Non-minimal constitutive equations}

The linear constitutive equation (\ref{eld2}) has the standard
form $H^{ik} = C^{ikmn} F_{mn}$ \cite{Maugin,Landau,HO} with a
``material" tensor
\begin{equation}
C^{ikmn} \equiv \frac{1}{2}(g^{im}g^{kn} - g^{in}g^{km}) + {\cal
R}^{ikmn}  \,. \label{ce1}
\end{equation}
This tensor describes the linear electromagnetic response of the
system, which may also be characterized by the dielectric and
magnetic permeabilities, as well as by possible magneto-electric
effects \cite{Maugin,Landau,HO}. $C^{ikmn}$ can uniquely be
decomposed with respect to the four velocity $U^i$ (normalized by
$U^iU_i=1$) of the medium:
\begin{eqnarray}
C^{ikmn} &=& \frac12 \left[ \varepsilon^{im} U^k U^n -
\varepsilon^{in} U^k U^m +
\varepsilon^{kn} U^i U^m - \varepsilon^{km} U^i U^n \right] \nonumber\\
&&-\frac12
\eta^{ikl}(\mu^{-1})_{ls}  \eta^{mns}  \nonumber\\
&& -\frac12 \left[\eta^{ikl}(U^m\nu_{l \ \cdot}^{\ n} - U^n \nu_{l
\ \cdot}^{\ m}) + \eta^{lmn}(U^i \nu_{l \ \cdot}^{\ k} - U^k
\nu_{l \ \cdot}^{\ i} ) \right] . \label{Cdecomp}
\end{eqnarray}
Here $\varepsilon^{im}$ and  $(\mu^{-1})_{pq}$ are the dielectric
and magnetic permeability tensors, respectively, and $\nu_{p \
\cdot}^{\ m}$  is a tensor of magneto-electric coefficients. These
quantities are defined as
\begin{equation}
\varepsilon^{im} = 2 C^{ikmn} U_k U_n, \quad (\mu^{-1})_{pq} = -
\frac{1}{2} \eta_{pik} C^{ikmn} \eta_{mnq} \,, \label{emu}
\end{equation}
\begin{equation}
\nu_{p \ \cdot}^{\ m} = \eta_{pik} C^{ikmn} U_n =U_k C^{mkln}
\eta_{lnp} \,. \label{nu}
\end{equation}
The dot denotes the position of the second index when lowered. The
tensors $\eta_{mnl}$ and $\eta^{ikl}$ are anti-symmetric and
orthogonal to $U^i$,
\begin{equation}
\eta_{mnl} \equiv \epsilon_{mnls} U^s \,, \quad \eta^{ikl} \equiv
\epsilon^{ikls} U_s \,. \label{eta}
\end{equation}
They satisfy the identity
\begin{equation}
- \eta^{ikp} \eta_{mnp} = \delta^{ikl}_{mns} U_l U^s = \Delta^i_m
\Delta^k_n - \Delta^i_n \Delta^k_m \,, \label{etaid1}
\end{equation}
where $\delta^{ikl}_{mns}$ is the generalized 6-indices $\delta-$
Kronecker tensor. The spatial projection tensor $\Delta^{ik}$ is
defined by
\begin{equation}
\Delta^{ik} = g^{ik} - U^i U^k \,. \label{projector}
\end{equation}
Contracting  equation (\ref{etaid1}) yields
\begin{equation}
\frac{1}{2} \eta^{ikl}  \eta_{klm} = - \delta^{il}_{ms} U_l U^s =
- \Delta^i_m \,. \label{etaid2}
\end{equation}
The tensors $\varepsilon_{ik}$ and $(\mu^{-1})_{ik}$ are
symmetric, but $\nu_{l k}$ is generally non-symmetric. All these
tensors are orthogonal to  $U^i$,
\begin{equation}
\varepsilon_{ik} U^k = 0, \quad (\mu^{-1})_{ik} U^k = 0, \quad
\nu_{l \ \cdot}^{\ k} U^l = 0 = \nu_{l \ \cdot}^{\ k} U_k \,.
\label{orto}
\end{equation}
Use of (\ref{ce1}) in (\ref{emu}) and (\ref{nu}) provides us with
\begin{equation}
\varepsilon^{im} = \Delta^{im} + 2 {\cal R}^{ikmn} U_k U_n \,,
\label{Re}
\end{equation}
\begin{equation}
(\mu^{-1})_{pq} = \Delta_{pq} - \frac{1}{2} \eta_{pik} {\cal
R}^{ikmn} \eta_{mnq} \,, \label{Rmu}
\end{equation}
\begin{equation}
\nu_{p \ \cdot}^{\ m} =  \eta_{pik} {\cal R}^{ikmn} U_n \,.
\label{Rnu}
\end{equation}
The non-minimal coupling of gravitational and electromagnetic
fields may be interpreted as a change of the dielectric and
magnetic properties of the vacuum, including a specific
magnetoelectric interaction. The vacuum acquires properties of a
quasi-medium under the influence of a non-vanishing tensor ${\cal
R}^{ikmn}$. The analogy between non-minimally extended
electrodynamics and macroscopic media was pointed out, e.g., in
\cite{CollKost}. This analogy may be completed by introducing the
electric induction $D^i$, the magnetic field $H^i$, the electric
field $E^i$ and the magnetic induction $B^i$, \cite{Maugin}:
\begin{equation}
D^i = \varepsilon^{im} E_m - B^l \nu_{l \ \cdot}^{\ i} \,, \quad
H_i = \nu_{i \ \cdot}^{\ m} E_m + (\mu^{-1})_{im} B^m \,.
\label{DH}
\end{equation}
The vectors $D^i$, $H^i$, $E^i$ and $B^i$ are defined by
\cite{Lichnero}:
\begin{equation}
D^i = H^{ik} U_k \,, \quad H^i = H^{*ik} U_k \,, \quad E^i =
F^{ik} U_k \,, \quad B^i = F^{*ik} U_k \,. \label{DHEB}
\end{equation}
They are orthogonal to the velocity four-vector $U^i$:
\begin{equation}
D^i U_i = 0 = E^i U_i \,, \quad H^i U_i = 0 = B^i U_i \,,
\label{orthogon}
\end{equation}
and form the basis for the decomposition of the tensors $F_{mn}$
and $H_{mn}$:
\begin{equation}
F_{mn} = E_m U_n - E_n U_m - \eta_{mnl} B^l \,, \quad H_{mn} = D_m
U_n - D_n U_m - \eta_{mnl} H^l \,. \label{FHdecomp}
\end{equation}

\section{Gravitational field equations}
\label{Gravitational field}

The gravitational field equations are obtained by varying the
action (\ref{act}) with the Lagrangian (\ref{lag}) with respect to
the metric tensor. They can be written in the standard form
\begin{equation}
R_{ik} - \frac{1}{2} R \ g_{ik} = \Lambda \ g_{ik} + \kappa
T^{({\rm eff})}_{ik} \,, \label{Ein}
\end{equation}
where
\begin{equation}
T^{({\rm eff})}_{ik} = T^{({\rm matter})}_{ik} + T^{(0)}_{ik} +
q_1 T^{(1)}_{ik} + q_2 T^{(2)}_{ik} + q_3 T^{(3)}_{ik} \,.
\label{Tdecomp}
\end{equation}
The stress-energy tensor of the matter $T^{({\rm matter})}_{ik}$
may be decomposed according to
\begin{equation}
T^{({\rm matter})}_{ik} = W U_i U_k  + I^{(q)}_i U_k + I^{(q)}_k
U_i + P_{ik}  \,, \label{Tmat}
\end{equation}
where $W$ is the matter energy density scalar, $P_{ik}$ is the
symmetric (generally anisotropic) pressure tensor, orthogonal to
the velocity four-vector ($P_{ik}U^k = 0$), and $I^{(q)}_i$ is the
energy-flux four-vector, orthogonal to the four velocity
($I^{(q)}_i U^i=0$). By $T^{(0)}_{ik}$ we denote the usual
stress-energy tensor of the electromagnetic field,
\begin{equation}
\quad T^{(0)}_{ik} \equiv \frac{1}{4} g_{ik} F_{mn}F^{mn} -
F_{in}F_{k}^{\cdot n} \,.   \label{T0}
\end{equation}
The contributions from the non-minimal interaction are
\begin{eqnarray}
T^{(1)}_{ik} = R \ T^{(0)}_{ik} - \frac{1}{2} R_{ik}
F_{mn}F^{mn}\qquad\qquad\qquad\qquad\qquad\nonumber\\ -
\frac{1}{2} g_{ik} \nabla^l \nabla_l (F_{mn}F^{mn}) + \frac{1}{2}
\nabla_{i} \nabla_{k} (F_{mn}F^{mn}) \,, \label{T1}
\end{eqnarray}
\begin{eqnarray}
T^{(2)}_{ik} = - \frac{1}{2}g_{ik}\left[ \nabla_{m}
\nabla_{l}(F^{mn}F^{l}_{\cdot n} ) - R_{lm}F^{mn}F^{l}_{\cdot
n}\right] - F^{ln}(R_{il}F_{kn} + R_{kl}F_{in}) \nonumber\\ -
R^{mn}F_{im}F_{kn} - \frac{1}{2} \nabla^l \nabla_l
(F_{in}F_{k}^{\cdot n}) \qquad\qquad\qquad\qquad\qquad
\nonumber\\+ \frac{1}{2}\nabla_l \left[ \nabla_i(F_{kn}F^{ln}) +
\nabla_k(F_{in}F^{ln}) \right] \,,\qquad\qquad\qquad \label{T2}
\end{eqnarray}
and
\begin{eqnarray}
T^{(3)}_{ik} = \frac{1}{4}g_{ik} R^{mnls}F_{mn}F_{ls} {-}
\frac{3}{4}F^{ls}(F_{i}^{\cdot n}R_{knls}+F_{k}^{\cdot n}R_{inls})
\qquad\qquad\nonumber\\{-} \frac{1}{2}\nabla_{m}
\nabla_{n}(F_{i}^{\cdot n}F_{k}^{\cdot m} {+} F_{k}^{\cdot
n}F_{i}^{\cdot m})\,.\qquad\qquad\qquad\label{T3}
\end{eqnarray}
The tensor $T^{(1)}_{ik}$ is proportional to the corresponding
term in \cite{Novello3}, the part $T^{(3)}_{ik}$ reproduces the
stress-energy tensor of \cite{Acci3}. The tensor $T^{(2)}_{ik}$ is
new. All three terms are supposed to contribute to the total
stress-energy tensor in the following. In contrast to the
traceless electromagnetic stress-energy tensor $T^{(0)}_{ik}$ the
tensors $T^{(1)}_{ik}$, $T^{(2)}_{ik}$ and $T^{(3)}_{ik}$ have
non-vanishing traces:
\begin{eqnarray}
g^{ik} T^{(1)}_{ik} &=&  - q_1 \left[ \frac{1}{2} R F_{mn}F^{mn} +
\frac{3}{2} \nabla^{k} \nabla_{k} (F_{mn}F^{mn}) \right] \,,\label{Tspur1}\\
g^{ik} T^{(2)}_{ik} &=& - q_2 \left[ R^{mn}F^{k}_{\cdot m}F_{kn} +
\frac{1}{2} \nabla^k \nabla_k (F_{mn}F^{mn})  \right] \,,\label{Tspur2}\\
g^{ik} T^{(3)}_{ik}  &=& - q_3 \left[ \frac{1}{2}
R^{mnls}F_{mn}F_{ls} + \nabla^{m} \nabla_{n}(F^{kn}F_{km}) \right]
\,. \label{Tspur}
\end{eqnarray}
Non-vanishing traces of effective stress energy tensors are also
features of non-linear electrodynamic models (see, e.g.,
\cite{LemosKerner}).

The effective stress-energy tensor (\ref{Tdecomp}) in
eq~(\ref{Ein}) has to be divergence-free, i.e.
\begin{equation}
\nabla^k T^{({\rm eff})}_{ik} = 0 \,. \label{nablaT}
\end{equation}
We assume the stress-energy tensor of the matter $T^{({\rm
matter})}_{ik}$ to be conserved separately, i.e., $ \nabla^k
T^{({\rm matter})}_{ik} = 0 $.  The remaining part of the
effective stress-energy tensor is then automatically conserved if
$F_{ik}$ is a solution of Maxwell's equations. In order to check
this fact directly, one has to use the Maxwell equations
(\ref{eld1}) with (\ref{eld2}), the Bianchi identities, the
symmetry properties of the Riemann tensor and the commutation
rules for the covariant derivatives. This procedure is analogous
to the one described in \cite{Acci3} and we omit it.

\section{Anisotropic cosmological models}
\label{Anisotropic models}

\subsection{Metric structure}

We consider now the Bianchi I cosmological model with the line
element \cite{ExactSolutions,ZelNov}
\begin{equation}
ds^2 = dt^2 - a^2(t) \ (dx^1)^2 - b^2(t) \ (dx^2)^2 - c^2(t) \
(dx^3)^2 \,. \label{metric}
\end{equation}
Due to the symmetry of the metric only six components of the
Riemann tensor are different from  zero:
$$
R^{01}_{\cdot \cdot 01} = - \frac{\ddot{a}}{a} \,, \quad
R^{02}_{\cdot \cdot 02} = - \frac{\ddot{b}}{b} \,, \quad
R^{03}_{\cdot \cdot 03} = - \frac{\ddot{c}}{c} \,, \quad
$$
\begin{equation}
R^{12}_{\cdot \cdot 12} = - \frac{\dot{a}}{a}\frac{\dot{b}}{b} \,,
\quad  R^{13}_{\cdot \cdot 13} = -
\frac{\dot{a}}{a}\frac{\dot{c}}{c} \,, \quad R^{23}_{\cdot \cdot
23} = - \frac{\dot{b}}{b} \frac{\dot{c}}{c} \,. \label{Raniso}
\end{equation}

\subsection{Exact solution of Maxwell's equations}

Since the susceptibility tensor ${\cal R}^{ikmn}$ has no
non-vanishing components with only one index zero (${\cal
R}^{\alpha \beta \gamma 0}=0$, Greek indices denote spatial
coordinates), it follows from relation (\ref{Rnu}) that all
magneto-electric coefficients vanish. The gravitational field
(\ref{metric}) does not mix pure electric and pure magnetic
fields. From relations (\ref{Re}) and (\ref{Rmu}) we find the
dielectric and magnetic permeability tensors
\begin{equation}
\varepsilon^{\alpha}_{\beta} = \delta^{\alpha}_{\beta} + 2 {\cal
R}^{\alpha}_{ 0 \beta 0}  \,, \quad (\mu^{-1})^{\alpha}_{\beta} =
\delta^{\alpha}_{\beta} - \frac{1}{2} \eta^{\alpha \gamma \sigma}
{\cal R}_{\gamma \sigma \cdot \cdot}^{ \ \ \mu \nu} \eta_{\mu \nu
\beta} \,. \label{emuaniso}
\end{equation}
Let us consider now a magnetic field directed along the $0z$ axis
(which is also the direction of the shear eigenvector
\cite{Maar}). The symmetry of the problem then fixes $F_{ik}$ and
$H^{ik}$ to be of the following structure:
\begin{equation}
F_{ik}  = (\delta^1_i \delta^2_k - \delta^2_i \delta^1_k) F_{12}
\,, \label{Fbian}
\end{equation}
\begin{equation}
H^{ik}  = (g^{i1}g^{k2}-g^{i2}g^{k1}) [1 + q_1 R + q_2 (R^1_1 +
R^2_2) + q_3 R^{12}_{\cdot \cdot 12}]F_{12} \,. \label{Hbian}
\end{equation}
The second set of the equations (\ref{eld1}) yields $F_{12}= {\rm
const}$. The first set of eqs.~(\ref{eld1}) is identically
satisfied since all the components $H^{i0}$ are equal to zero.
Introducing the scalar value of the magnetic field $B(t)$ by
\begin{equation}
B^2(t) = \frac{1}{2} F_{ik} F^{ik} = F_{12} F^{12} \,,\label{BF2}
\end{equation}
we reproduce the result [cf. \cite{ZelNov})]
\begin{equation}
B^2(t)= \frac{{\rm const}^2}{a^2(t) b^2(t)} \,, \label{Binv}
\end{equation}
i.e., $B(t) a(t) b(t) = {\rm const}$.
\newpage

\subsection{Einstein's equations}

In a co-moving frame with $U^i=\delta^i_0$, the field equations
(\ref{Ein}) reduce to the following system:
\begin{eqnarray}
\frac{\dot{a}}{a} \frac{\dot{b}}{b} + \frac{\dot{a}}{a}
\frac{\dot{c}}{c} + \frac{\dot{b}}{b} \frac{\dot{c}}{c} &=&
\Lambda + \kappa W  +  \frac{1}{2} \kappa B^2(t)
\nonumber\\
&&{+} \kappa B^2(t)\left\{ q_1
\left[2\left(\left(\frac{\dot{a}}{a}\right)^2 {+}
\left(\frac{\dot{b}}{b}\right)^2\right) {+} 3 \frac{\dot{a}}{a}
\frac{\dot{b}}{b} {+} \frac{\dot{c}}{c}\left(\frac{\dot{a}}{a} {+}
\frac{\dot{b}}{b} \right) \right] \right.\nonumber\\ &&
\qquad\qquad\qquad\qquad + \left. q_2 \left(\frac{\dot{a}}{a} {+}
\frac{\dot{b}}{b} \right)^2 {+} q_3 \frac{\dot{a}}{a}
\frac{\dot{b}}{b} \right\} \,,\label{Ein00}
\end{eqnarray}
\begin{eqnarray}
\frac{\ddot{b}}{b} + \frac{\ddot{c}}{c} + \frac{\dot{b}}{b}
\frac{\dot{c}}{c} = \Lambda - \kappa P_{(1)}  - \frac{1}{2} \kappa
B^2(t)\qquad\qquad\qquad\qquad\qquad\qquad\qquad\qquad\nonumber\\
+ \kappa B^2(t)\left\{ q_1 \left[ 4 \frac{\ddot{a}}{a} + 3
\frac{\ddot{b}}{b} + \frac{\ddot{c}}{c}  - 6
\left(\frac{\dot{a}}{a}\right)^2 - 4
\left(\frac{\dot{b}}{b}\right)^2 - 4 \frac{\dot{a}}{a}
\frac{\dot{b}}{b} + 4 \frac{\dot{a}}{a} \frac{\dot{c}}{c} + 3
\frac{\dot{b}}{b} \frac{\dot{c}}{c} \right] \right. \nonumber\\
+ q_2 \left[ 2 \left(\frac{\ddot{a}}{a} {+} \frac{\ddot{b}}{b}
\right) {-} 3 \left(\left(\frac{\dot{a}}{a}\right)^2 {+}
\left(\frac{\dot{b}}{b}\right)^2\right) {-} 2 \frac{\dot{a}}{a}
\frac{\dot{b}}{b} {+} 2 \frac{\dot{c}}{c}\left(\frac{\dot{a}}{a}
{+} \frac{\dot{b}}{b} \right) \right] \nonumber\\{+}\left. q_3
\left[\frac{\ddot{b}}{b} {-} 2 \left(\frac{\dot{b}}{b}\right)^2
{+} \frac{\dot{b}}{b} \frac{\dot{c}}{c} \right] \right \} \,,
\label{Ein11}
\end{eqnarray}
\begin{eqnarray}
\frac{\ddot{a}}{a} + \frac{\ddot{c}}{c} + \frac{\dot{a}}{a}
\frac{\dot{c}}{c} = \Lambda - \kappa P_{(2)} -
\frac{1}{2} \kappa B^2(t) \qquad\qquad\qquad\qquad\qquad\qquad\qquad\qquad
\nonumber\\
+ \kappa B^2(t)\left\{ q_1 \left[ 4 \frac{\ddot{b}}{b} + 3
\frac{\ddot{a}}{a} + \frac{\ddot{c}}{c}  - 6
\left(\frac{\dot{b}}{b}\right)^2 - 4
\left(\frac{\dot{a}}{a}\right)^2 - 4 \frac{\dot{a}}{a}
\frac{\dot{b}}{b} + 4 \frac{\dot{b}}{b} \frac{\dot{c}}{c} + 3
\frac{\dot{a}}{a} \frac{\dot{c}}{c} \right]  \right.\nonumber\\
{+} q_2 \left[ 2 \left(\frac{\ddot{a}}{a} {+} \frac{\ddot{b}}{b}
\right) {-} 3 \left(\left(\frac{\dot{a}}{a}\right)^2 {+}
\left(\frac{\dot{b}}{b}\right)^2\right) {-} 2 \frac{\dot{a}}{a}
\frac{\dot{b}}{b} {+} 2 \frac{\dot{c}}{c}\left(\frac{\dot{a}}{a}
{+} \frac{\dot{b}}{b} \right) \right] \nonumber\\ {+}\left. q_3
\left[\frac{\ddot{a}}{a} {-} 2 \left(\frac{\dot{a}}{a}\right)^2
{+} \frac{\dot{a}}{a} \frac{\dot{c}}{c} \right] \right\} \,,
\label{Ein22}
\end{eqnarray}
\newpage
\begin{eqnarray}
\frac{\ddot{a}}{a} + \frac{\ddot{b}}{b} + \frac{\dot{a}}{a}
\frac{\dot{b}}{b} = \Lambda - \kappa P_{(3)} + \frac{1}{2} \kappa
B^2(t)
\qquad\qquad\qquad\qquad\qquad\qquad\qquad\qquad\nonumber\\ +
\kappa B^2(t)\left\{ q_1 \left[\frac{\ddot{a}}{a} +
\frac{\ddot{b}}{b} - 4 \left( \left(\frac{\dot{a}}{a}\right)^2 +
\left(\frac{\dot{b}}{b}\right)^2 \right) - 5 \frac{\dot{a}}{a}
\frac{\dot{b}}{b} \right] \right.\nonumber\\- \left. q_2
\left(\frac{\dot{a}}{a} + \frac{\dot{b}}{b} \right)^2 - q_3
\frac{\dot{a}}{a} \frac{\dot{b}}{b} \right\} \,. \label{Ein33}
\end{eqnarray}
Here, $P_{(1)}$, $P_{(2)}$ and $P_{(3)}$ are the eigenvalues of
the anisotropic pressure tensor $P_{ik}$. Summing up
eqs.~(\ref{Ein00})-(\ref{Ein33}) we obtain the trace equation:
\begin{eqnarray}
2 \left( \frac{\ddot{a}}{a} + \frac{\ddot{b}}{b} +
\frac{\ddot{c}}{c}+ \frac{\dot{a}}{a} \frac{\dot{b}}{b} +
\frac{\dot{a}}{a} \frac{\dot{c}}{c} + \frac{\dot{b}}{b}
\frac{\dot{c}}{c} \right) = 4 \Lambda + \kappa ( W - P_{(1)} -
P_{(2)} - P_{(3)}) \nonumber\\
+ \kappa B^2(t)\left\{ (8 q_1 + 4 q_2 + q_3)
\left(\frac{\ddot{a}}{a} + \frac{\ddot{b}}{b} \right)  + 2q_1
\frac{\ddot{c}}{c} \right.\qquad\qquad\qquad\qquad\nonumber\\ - 2
(6 q_1 + 3 q_2 + q_3) \left( \left(\frac{\dot{a}}{a}\right)^2 +
\left(\frac{\dot{b}}{b}\right)^2 \right)  - 2 (5 q_1 + 2 q_2)
\frac{\dot{a}}{a} \frac{\dot{b}}{b} \nonumber\\
\left. + (8 q_1 + 4 q_2 + q_3) \frac{\dot{c}}{c}
\left(\frac{\dot{a}}{a} + \frac{\dot{b}}{b} \right) \right\}\,.
\label{Eintrace}
\end{eqnarray}
\\
Differentiating (\ref{Ein00}) and using (\ref{Ein11}) -
(\ref{Ein33}) leads  to the conservation law for the matter:
\begin{equation}
\dot{W} + \left( \frac{\dot{a}}{a} + \frac{\dot{b}}{b} +
\frac{\dot{c}}{c}\right) W +  \frac{\dot{a}}{a} P_{(1)} +
\frac{\dot{b}}{b} P_{(2)} + \frac{\dot{c}}{c} P_{(3)}  = 0 \,.
\label{conserva}
\end{equation}
It follows that $T^{(0)}_{ik} + q_1 T^{(1)}_{ik} + q_2
T^{(2)}_{ik} + q_{3} T^{(3)}_{ik}$ is conserved as well. The
equations (\ref{Ein00}) - (\ref{Ein33}) represent a modified
(compared to minimal coupling) dynamical system, since the
non-minimal terms contribute to the coefficients before the second
order derivatives. Its complete analysis should be the subject of
future investigations along the lines described in \cite{dina}.

\section{Particular models}
\label{Particular models}

\subsection{Quasi-one-dimensional solutions with pure magnetic field}

As the first application we consider a model without matter
($W=0$, $P_{ik}= 0$) and with constant values of two of the metric
functions: $a(t)= a_0$ and $b(t)= b_0$. Only $c(t)$ is assumed to
vary. Then $B^2(t) = {\rm const}$ (see, (\ref{Binv})) and the
equations (\ref{Ein00})-(\ref{Ein33}) reduce to
\begin{equation}
0 = \Lambda + \frac{1}{2}\kappa B^2(t_0) \,, \quad
\frac{\ddot{c}}{c} \left[1 - q_1 \kappa B^2(t_0) \right] = \Lambda
- \frac{1}{2}\kappa B^2(t_0) \,. \label{long1}
\end{equation}
Only one of the non-minimal parameters enters the dynamics.
According to the first equation this model requires a negative
cosmological constant which exactly compensates the magnetic field
term, i.e., $\kappa B^2 = - 2 \Lambda$. The equation for $c(t)$ is
\begin{equation}
\ddot{c}(t) + c(t) \left[\frac{\kappa B^2(t_0)}{1 - q_1 \kappa
B^2(t_0)} \right] = 0 \,. \label{long2}
\end{equation}
For $ B^2 = \Lambda = 0$ we obtain the vacuum solution $c \propto
t$ which is the degenerate case of a Kasner solution, equivalent
to the flat spacetime Milne universe (cf. \cite{ZelNov}). For
non-vanishing $\kappa B^2 = - 2 \Lambda$ we have three cases.

\vspace{3mm} \noindent {\it First case}: $q_1 \kappa B^2(t_0) <
1$.

\noindent This condition includes the case $q_1=0$. The solution
of eq.~(\ref{long2}) is oscillatory:
\begin{equation}
c(t) =  c(t_0) \cos \nu (t-t_0) + \frac{\dot{c}(t_0)}{\nu} \sin
\nu (t-t_0) \,, \label{osc}
\end{equation}
where
\begin{equation}
\nu^2 \equiv \left[ \frac{\kappa B^2(t_0)}{1 - q_1 \kappa
B^2(t_0)} \right] \,. \label{long3}
\end{equation}
The zeros of $c(t)$ denote singularities of the model.

\vspace{3mm} \noindent {\it Second case}: $q_1 \kappa B^2(t_0)
=1$.

\noindent The equations (\ref{long1}) contradict each other. This
case is incompatible with $B(t_0) \neq 0$.

\vspace{3mm} \noindent {\it Third case}: $q_1 \kappa B^2(t_0) >
1$.

\noindent This condition requires $q_1$ to be positive. The
solution of (\ref{long2}) is
\begin{equation}
c(t) =  c(t_0) \cosh \mu (t-t_0) + \frac{\dot{c}(t_0)}{\mu} \sinh
\mu (t-t_0) \,, \label{exp}
\end{equation}
where
\begin{equation}
\mu^2 \equiv \left[ \frac{\kappa B^2(t_0)}{q_1 \kappa B^2(t_0) -
1} \right] \,. \label{long4}
\end{equation}
For $t >> t_0$ the function $c(t)$ behaves as $c(t) \propto e^{\mu
t}$. The model is non-singular when $|\frac{c(t_0)
\mu}{\dot{c}(t_0)}| \geq 1$. If $|\frac{c(t_0) \mu}{\dot{c}(t_0)}|
< 1$, then there exists a singularity at $t^{*}$, given by $\tanh
\mu (t^{*}-t_0) = - \frac{c(t_0) \mu}{\dot{c}(t_0)}$ and
$c(t^{*})= 0$.

\subsection{Quasi-two-dimensional solutions with magnetic field and matter}

A second model with $B(t)= $const is obtained for $a(t)b(t)=
$const. We may write $a(t) = a(t_0)E(t)$ and $b(t) = b(t_0)
E^{-1}(t)$ with $E(t_0)= 1$ where $t_0$ is some reference time. If
we additionally assume $c(t) =$ const, the dynamics is restricted
to the $x^1Ox^2$ plane. The equations (\ref{Ein00})-(\ref{Ein33})
reduce to the system:
\begin{equation}
- L \left( \frac{\dot{E}}{E}\right)^2  = \Lambda + \kappa W +
\frac{1}{2} \kappa B^2 \,, \quad - L \frac{\ddot{E}}{E}  + 2
\left( \frac{\dot{E}}{E}\right)^2 = \Lambda - \kappa P_{(1)} -
\frac{1}{2} \kappa B^2 \,,\label{trav01}
\end{equation}
\\
\begin{equation}
L \left( \frac{\dot{E}}{E}\right)^2  = \Lambda - \kappa P_{(3)} +
\frac{1}{2} \kappa B^2 \,, \quad L \frac{\ddot{E}}{E}  + 2 (1 - L)
\left( \frac{\dot{E}}{E}\right)^2  = \Lambda - \kappa P_{(2)} -
\frac{1}{2} \kappa B^2 \,, \label{trav1}
\end{equation}
\\
where
\begin{equation}
L \equiv 1 + (q_1 - q_3) \kappa B^2 = {\rm const} \,. \label{L}
\end{equation}
\\
For $L \neq 0$ the system (\ref{trav01}), (\ref{trav1}) is
equivalent to
\begin{equation}
2L \left( \frac{\dot{E}}{E}\right)^2  = - \kappa ( W + P_{(3)})
\,, \quad 2L  \left( \frac{\dot{E}}{E} \right)^{\cdot} =  \kappa (
P_{(1)} - P_{(2)} ) \,,\label{trav02}
\end{equation}
\\
\begin{equation}
P_{(3)} = W + B^2 + \frac{2 \Lambda}{\kappa} \,, \quad P_{(1)} +
P_{(2)} = 2 W +  \frac{4 \Lambda}{\kappa} \,. \label{trav2}
\end{equation}
\\
The minimally coupled case $q_1 = q_2 = q_3 = 0$ corresponds to
$L=1$. It is dynamically indistinguishable from a non-minimal
configuration with $q_1 = q_3$ and arbitrary $q_2$. The case $L=1$
requires $W + P_{(3)} <0$. This can only be achieved by a
sufficiently negative cosmological constant, $2W + B^2 + \frac{2
\Lambda}{\kappa} <0 $ which also implies that $P_{(1)} + P_{(2)} <
0$.

\subsubsection{Case $L=0$}

For $q_3 - q_1 = \frac{1}{\kappa B^2}$ the quantity $E(t)$ is
constant, i.e., the universe is static. The matter distribution is
characterized by constant quantities as well:
\begin{equation}
P_{(1)} = P_{(2)} =  \frac{\Lambda}{\kappa} - \frac{1}{2} B^2 \,,
\quad  W = - P_{(3)} = - \frac{\Lambda}{\kappa} - \frac{1}{2}
B^2\,. \label{zeroL}
\end{equation}
The energy density $W$ is positive if $\Lambda < - \frac{1}{2}
\kappa B^2$. All the pressure eigenvalues become negative in this
case.

\subsubsection{Case $L\neq 0$, $P_{(1)} = P_{(2)} $}

Here we obtain
\begin{equation}
E(t) = e^{H_0(t-t_0)} \,, \quad  W = - L H^2_0 -
\frac{\Lambda}{\kappa} - \frac{1}{2} B^2 \,,\label{sitter0}
\end{equation}
\begin{equation}
P_{(3)} = - L H^2_0 + \frac{\Lambda}{\kappa} + \frac{1}{2} B^2 \,,
\quad P_{(1)} = P_{(2)} = - L H^2_0 + \frac{\Lambda}{\kappa} -
\frac{1}{2} B^2 \,, \label{sitter1}
\end{equation}
where $H_0$ is an arbitrary integration constant. For any $L
> 0$ a sufficiently negative cosmological constant is required for the
energy density $W$ to be positive. After re-parametrization of the
coordinates and the time the metric takes a form
\begin{equation}
ds^2 = dt^2 - \left( e^{H_0 t} dx^2 +  e^{- H_0 t} dy^2 \right) -
dz^2 \,. \label{sitter2}
\end{equation}

\subsubsection{Ultrarelativistic matter with $L\neq
0$, $P_{(1)} = P_{(2)}$}

If the matter is ultrarelativistic, i.e., $T^k_{k ({\rm
matter})}=0$ and, consequently, $W = P_{(1)} + P_{(2)} + P_{(3)}$,
one obtains
\begin{equation}
H^2_0 = \frac{2\Lambda}{\kappa L} \,, \quad W = - \frac{3}{2}L
H^2_0 -   \frac{1}{2} B^2 \,,\label{sitter03}
\end{equation}
\begin{equation}
P_{(3)} = - \frac{1}{2}L H^2_0 + \frac{1}{2} B^2 \,, \quad P_{(1)}
= P_{(2)} = - \frac{1}{2}L H^2_0 - \frac{1}{2} B^2 \,.
\label{sitter3}
\end{equation}
The energy density $W$ is positive for $LH^2_0 < - \frac{1}{3}
B^2$, which requires the constant $L$ to be negative, i.e., $q_3 -
q_1 > \frac{1}{\kappa B^2}$. In this case the longitudinal
pressure $P_{(3)}$ is positive as well. The transversal pressure
$P_{(1)} = P_{(2)}$ is positive for $LH^2_0 < - B^2$. The
condition $L<0$ necessarily implies a negative cosmological
constant again.

\subsection{Axial symmetry: $a(t)=b(t)$, $P_{(1)} = P_{(2)} \equiv P_{({\rm tr})}$}

All the metric functions are assumed to be time dependent now.
Einstein's equations reduce to the following system of three
equations for two unknown functions:
\begin{equation}
\left(\frac{\dot{a}}{a}\right)^2 + 2 \frac{\dot{a}}{a}
\frac{\dot{c}}{c} = \Lambda + \kappa W +  \kappa B^2(t)\left\{
\frac{1}{2} + ( 7 q_1 + 4 q_2 + q_3)
\left(\frac{\dot{a}}{a}\right)^2 + 2 q_1
\frac{\dot{a}}{a}\frac{\dot{c}}{c} \right\} \,, \label{uni1}
\end{equation}
\begin{equation}
\frac{\ddot{a}}{a} {+} \frac{\ddot{c}}{c} {+} \frac{\dot{a}}{a}
\frac{\dot{c}}{c} {=} \Lambda {-} \kappa P_{({\rm tr})} {+} \kappa
B^2(t)\left\{{-} \frac{1}{2} {+} q_1 \frac{\ddot{c}}{c} {+} (7 q_1
{+} 4 q_2 {+} q_3) \left[ \frac{\ddot{a}}{a} {-} 2
\left(\frac{\dot{a}}{a}\right)^2 {+} \frac{\dot{a}}{a}
\frac{\dot{c}}{c} \right] \right\}  \,, \label{uni2}
\end{equation}
\begin{equation}
2 \frac{\ddot{a}}{a} + \left(\frac{\dot{a}}{a} \right)^2 = \Lambda
-  \kappa P_{(3)} + \kappa B^2(t)\left\{ \frac{1}{2} + 2 q_1
\frac{\ddot{a}}{a} - \left(\frac{\dot{a}}{a}\right)^2 (13 q_1 + 4
q_2 + q_3) \right\} \,. \label{uni3}
\end{equation}
\\
The matter conservation law takes the form
\begin{equation}
\dot{W} + 2 \frac{\dot{a}}{a} \left( W + P_{({\rm tr})} \right) +
\frac{\dot{c}}{c} \left( W + P_{(3)} \right) = 0 \,.
\label{uniconserva}
\end{equation}
\\
Formally, the function $c(t)$ does not appear in equation
(\ref{uni3}). However, in general there is a coupling to $c(t)$
via the pressure $P_{(3)}$ through the conservation law
(\ref{uniconserva}). There are two simple cases for which
eq.~(\ref{uni3}) decouples from eqs.~(\ref{uni1}) and
(\ref{uni2}). The first one is $P_{(3)}=0$, i.e., a vanishing
longitudinal pressure, the second one is $W + P_{(3)}=0$, i.e., a
vacuum type behavior in longitudinal direction together with a
transversal equation of state $P_{({\rm tr})} = P_{({\rm
tr})}(W)$. In the following we consider both cases separately.

\subsubsection{``Longitudinal dust": $P_{(3)}=0$}

With the substitution
\begin{equation}
\dot{a}(t) = a^{\sigma} \sqrt{Z(a)} \,, \quad \sigma = \frac{13
q_1 + 4 q_2 + q_3}{2 q_1} \,, \quad B^2(t) = \frac{M^2}{a^4} \,,
\label{substit}
\end{equation}
where $M^2 = $const and $q_1 \neq 0$, equation (\ref{uni3}) is
transformed into the first order equation
\begin{equation}
\left(1 - \frac{\kappa q_1 M^2}{a^4} \right) \frac{d Z(a)}{da} +
(2 \sigma + 1)\frac{Z}{a} = \left(\Lambda + \frac{\kappa M^2}{2
a^4} \right)a^{1-2 \sigma} \,. \label{Zeq}
\end{equation}
The solution of eq.~(\ref{Zeq}) can be represented as quadrature:
\begin{eqnarray}
Z(a) = H_a^2(a_0) a^{2 - 2 \sigma}_0 \left( \frac{a^4_0 - \kappa
q_1 M^2}{a^4 - \kappa q_1 M^2} \right)^{\frac{2 \sigma +1}{4}}
\qquad\qquad\qquad\qquad\qquad\qquad\qquad
\nonumber\\
+ \left( a^4 - \kappa q_1 M^2 \right)^{ - \frac{2 \sigma +1}{4}}
\int^a_{a_0} dx x^{1 - 2 \sigma} \left( x^4 - \kappa q_1 M^2
\right)^{  \frac{2 \sigma - 3}{4}} \left( \Lambda x^4 +
\frac{1}{2} \kappa M^2 \right) \,, \nonumber\\ \label{Zsol}
\end{eqnarray}
where $H_a(a_0) \equiv a^{\sigma -1} \sqrt{Z(a_0)}$ with $a_0
\equiv a(t_0)$ and $H_a(t) \equiv \frac{\dot{a}}{a} = H_b(t)
\equiv \frac{\dot{b}}{b} $ is the expansion rate in the $x^{1}$
and $x^{2}$ directions. Due to the expressions $ \left( x^4 -
\kappa q_1 M^2 \right)^{ \frac{2 \sigma - 3}{4}}$ and $ \left( a^4
- \kappa q_1 M^2 \right)^{ \frac{2 \sigma - 3}{4}}$ in
Eq.~(\ref{Zsol}), the result of the integration is sensitive to
the sign of $q_1$. For $q_1 > 0$ the term $ \left( x^4 - \kappa
q_1 M^2 \right)$ has two real zeros $x_{1,2}  = \pm (\kappa q_1
M^2)^{\frac{1}{4}}$. If at least one zero belongs to the interval
$(a_0, a(t))$, the integral in eq.~(\ref{Zsol}) diverges for
$\frac{2\sigma -3}{4} \leq -1$, i.e., for $2 \sigma + 1 \leq 0$.
If $q_1 < 0$, such a singularity does not appear. The asymptotic
behavior of the function $Z(a)$ for $a \to \infty$ is
\begin{equation}
Z(a \to \infty) = \frac{\Lambda}{3} a^{2 - 2 \sigma} \,.
\label{asymp1}
\end{equation}
Thus,  asymptotically all the models yield
\begin{equation}
\frac{\dot{a}}{a} = \sqrt{\frac{\Lambda}{3}} = H_a = {\rm const}
\,, \quad a(t) = a(t_0) e^{H_a (t-t_0)} \,, \label{asymp2}
\end{equation}
i.e., a de Sitter type expansion for $a(t)$, independent of the
parameter $\sigma$. Moreover, one can check directly that the
solution (\ref{asymp2}) is an exact solution of the equation
(\ref{uni3}) with
\begin{equation}
H_a^{-2} = 2 (11 q_1 + 4 q_2 + q_3) \,, \label{Hqqq}
\end{equation}
where $11 q_1 + 4 q_2 + q_3 \neq 0$. It is remarkable, that the
constant expansion rate $H_a$ is determined both by $\Lambda$ via
eq.~(\ref{asymp2}) and by the non-minimal coupling parameters via
eq.~(\ref{Hqqq}). The relation (\ref{Hqqq}) has no counterpart in
the minimal theory. Combining (\ref{asymp2}) and (\ref{Hqqq})
allows us to establish the following relation between $\Lambda$
and the coupling parameters:
\begin{equation}
\Lambda = \frac{3}{2\left(11 q_1 + 4 q_2 + q_3\right)}  \,.
\label{Lambdaq}
\end{equation}
The cosmological constant is expressed in terms of quantities
which are supposed to be the result of quantum field theoretical
calculations. In quantum electrodynamics the parameters $q_1$,
$q_2$, and $q_3$ are \cite{Drum} $q_1 = - \frac{\alpha
\lambda^2_e}{180 \pi}$, $q_2 = -13 q_1$, $q_3 = 2 q_1$, where
$\alpha$ is the fine structure constant, i.e., they are
proportional to the square of the Compton wavelength $\lambda_e$
of the electron. This would give rise to a value $\Lambda_{\rm
QED} = \frac{90}{13}\frac{\pi}{\alpha\lambda^2_e} $ of the
cosmological constant. While one does not expect a quantum
electrodynamical length to set the scale for an early de Sitter
stage (recall that we assume $q_1$, $q_2$, and $q_3$ to be free
parameters) this result may nevertheless indicate a potential
relevance of non-minimal interactions for an early inflationary
dynamics.

With a transversal equation of state $P_{({\rm tr})}(t) = (\gamma
- 1) W(t)$ and with (\ref{asymp2}) the conservation law
(\ref{uniconserva}) yields
\begin{equation}
W(t) = W(t_0) \frac{c(t_0)}{c(t)} e^{-2H_a \gamma (t-t_0)} \,.
\label{W}
\end{equation}
\\
The equation (\ref{uni1}) for $c(t)$ can be rewritten in the form
\begin{equation}
\frac{d}{d z} \left\{ [1 - \alpha z]^{- \xi} Y(z)\right\} = - [1 -
\alpha z]^{- (\xi + 1)} \frac{\kappa W(t_0)}{8 H^2_a}
z^{\frac{2\gamma -3}{4}} \,, \label{Y}
\end{equation}
where
\begin{equation}
z \equiv \left(\frac{a_0}{a(t)}\right)^4 \,, \quad \alpha \equiv
\frac{q_1 \kappa M^2}{a^4_0} \,, \quad \xi \equiv \frac{1 - 2q_1
H^2_a}{8 q_1 H^2_a} \,, \quad Y \equiv
\frac{c(t)a(t_0)}{a(t)c(t_0)} \,. \label{Y1}
\end{equation}
For $\alpha \neq 1$ the solution for $c(t)$ with the initial value
$c(t_0)$ has the following explicit form
\begin{eqnarray}
c(t) = c(t_0) e^{H_a (t-t_0)} \left[ 1 - \alpha
e^{-4 H_a (t-t_0)} \right]^{\xi}\cdot \qquad\qquad\qquad\qquad\qquad\nonumber\\
\cdot\left\{ \left( 1 - \alpha \right)^{{-}\xi} {-} \frac{\kappa
W(t_0)}{8 H^2_a} \int^{e^{- 4H_a(t-t_0)}}_1 dx x^{\frac{2\gamma
-3}{4}} \left[ 1 {-} \alpha x \right]^{ - (\xi + 1)} \right\} \,.
\label{c}
\end{eqnarray}
At $\alpha x =1$ the integral in eq.~(\ref{c}) is non-singular for
$\xi < 0$. In the asymptotic regime $t \to \infty$ one obtains
from (\ref{c})
\begin{equation}
c(t) \approx c(t_0) e^{H_a (t-t_0)} \Gamma \,, \label{c1}
\end{equation}
where the constant value $\Gamma$ is equal to
\begin{equation}
\Gamma \equiv \left( 1 - \alpha  \right)^{{-}\xi} - \frac{\kappa
W(t_0)}{8 H^2_a} \int^0_1 dx x^{\frac{2\gamma -3}{4}} \left[ 1 -
\alpha x \right]^{- (\xi + 1)} \label{Gamma} \,.
\end{equation}
For $x \to 0$ the constant $\Gamma$ remains finite for
$\frac{2\gamma -3}{4}
> -1$, i.e., $2\gamma + 1 > 0$.

The expression (\ref{c1}) shows that the expansion rate in $0x^3$
direction tends asymptotically to the expansion rate in the
orthogonal direction, i.e., the universe becomes isotropic. The
isotropization rate is characterized by the function $K(t)$,
\begin{equation}
K(t) \equiv \log{Y} \,, \quad \dot{K}(t) = \frac{\dot{c}(t)}{c(t)}
- \frac{\dot{a}(t)}{a(t)} \,,\quad  \dot{K}(t \to \infty) \to 0
\,. \label{K}
\end{equation}
\\
Since $11 q_1 + 4 q_2 + q_3 \neq 0$ was assumed in
Eq.~(\ref{Hqqq}), the minimal limit $q_1 = q_2 = q_3 = 0$ cannot
be taken here, since it implies $H_a \to \infty$. To check whether
or not a corresponding isotropization takes place in the minimally
coupled theory as well one has to solve the system (\ref{uni1}) -
(\ref{uni3}) with $q_1 = q_2 = q_3 = 0$. It is straightforward to
realize that a solution $3H_a^2 = \Lambda$ of the minimally
coupled system requires $P_{(3)} = \frac{1}{2}B^{2}$.
Consequently, a dust equation of state $P_{(3)} = 0$ means the
absence of the magnetic field. While the non-minimal theory in our
example admits an isotropization in the presence of a magnetic
field, the minimally coupled theory does not.

For  $\alpha =1$ the initial value problem for $c(t)$ with the
initial value $c(t_0)$ degenerates. This requires a special
treatment and we do not consider this case here.

\subsubsection{``Longitudinal quasi-vacuum": $W + P_{(3)}=0$}

When $P_{({\rm tr})}(t) = (\gamma - 1) W(t)$ and $W + P_{(3)}=0$,
the conservation law yields
\begin{equation}
W(t) = W(t_0) \left( \frac{a(t_0)}{a(t)} \right)^{2 \gamma} \,,
\label{longvac1}
\end{equation}
and $a(t)$ can be found in quadratures from the equation
(\ref{uni3}). The solution $a(t)= a(t_0) e^{H_a (t-t_0)}$  holds
in this case as well if the magnetic field term with $B^2(t)
\propto a^{-4}(t)$ is compensated by the longitudinal pressure
$P_{(3)}(t) = - W(t) \propto a^{- 2\gamma}(t)$. This requires a
transversal stiff matter equation of state, i.e., $\gamma = 2$. As
in the previous subsection (see the discussion following eq.
(\ref{Lambdaq})) we obtain two expressions for $H_a$,
\begin{equation}
H_a = \sqrt{\frac{\Lambda}{3}} \, \quad{\rm and} \quad H^2_a =
\frac{2 W(t_0)a^4_0 + M^2}{2 M^2\left(11 q_1 + 4 q_2 + q_3
\right)} \,,\label{longvac2}
\end{equation}
which again imply a relation between $\Lambda$ and the parameters
of the non-minimal interaction. From equation (\ref{uni1}) we find
\begin{equation}
c(t) = c(t_0) e^{H_a(t-t_0)} \left[ 1 - \alpha  e^{-4 H_a (t-t_0)}
\right]^{\zeta} \left[ 1 - \alpha \right]^{- \zeta} \,,
\label{longvac3}
\end{equation}
where
\begin{equation}
\zeta  \equiv \frac{2 W(t_0) a^4_0 + M^2 (1- 2q_1 H^2_a)}{8 q_1
M^2 H^2_a} \,. \label{longvac4}
\end{equation}
\\
Again, the solution for $c(t)$ has the same asymptotical behavior
as $a(t)$, i.e., the universe becomes isotropic. This solution is
non-singular if $q_1$ is negative. $c(t)$ increases monotonically
for $4\zeta <1$.  If $4 \zeta > 1$ the function $c(t)$ increases
after it passed a minimum value. If $q_1$ is positive, there
exists a time $t^{*}$ with  $c(t^{*})= 0$, where $t^{*}$ is given
by
\begin{equation}
t^{*} = t_0 + \frac{1}{4 H_a} \log{\alpha} \,,\qquad \alpha =
\frac{\kappa q_1 M^2}{a^4_0}\,. \label{longvac5}
\end{equation}
The model is then applicable for $t >t^{*}$.

Also in this case the isotropization process is a property of the
non-minimal theory only. The corresponding solution $H_a =$ const
of the minimal theory is necessarily isotropic for $W +
P_{(3)}=0$.

\subsection{Isotropic universe model with ``hidden" magnetic field}

Let us consider now the conditions, under which an isotropic model
with $a(t) = b(t) = c(t)$ is compatible with the existence of a
magnetic field. The Einstein equations reduce to
\begin{equation}
3 \left(\frac{\dot{a}}{a} \right)^2 = \Lambda + \kappa W + \kappa
B^2(t)\left\{ \frac{1}{2} + ( 9 q_1 + 4 q_2 + q_3)
\left(\frac{\dot{a}}{a}\right)^2 \right\} \,, \label{iso1}
\end{equation}
\\
\begin{equation}
2 \frac{\ddot{a}}{a}  {+} \left(\frac{\dot{a}}{a} \right)^2 {=}
\Lambda {-} \kappa P_{({\rm tr})}  {+} \kappa B^2(t)\left\{{-}
\frac{1}{2} {+} (8 q_1 {+} 4 q_2 {+} q_3) \frac{\ddot{a}}{a} {-}
\left(\frac{\dot{a}}{a}\right)^2 (7 q_1 {+} 4 q_2 {+} q_3)
\right\}  \label{iso2}
\end{equation}
and
\begin{equation}
2 \frac{\ddot{a}}{a} + \left(\frac{\dot{a}}{a} \right)^2 = \Lambda
- \kappa P_{(3)} + \kappa B^2(t)\left\{ \frac{1}{2} + 2 q_1
\frac{\ddot{a}}{a} - \left(\frac{\dot{a}}{a}\right)^2 (13 q_1 + 4
q_2 + q_3) \right\} \,. \label{iso3}
\end{equation}
Furthermore, the trace equation becomes
\begin{equation}
6 \left[ \frac{\ddot{a}}{a} +  \left( \frac{\dot{a}}{a} \right)^2
\right] = 4 \Lambda + \kappa ( W - 2 P_{({\rm tr})} - P_{(3)} ) +
2 \kappa B^2(t) (9 q_1 + 4 q_2 + q_3) \left[ \frac{\ddot{a}}{a} -
\left(\frac{\dot{a}}{a}\right)^2 \right] \,, \label{iso4}
\end{equation}
and the conservation law simplifies to
\begin{equation}
\dot{W} +  \frac{\dot{a}}{a} \left( 3 W  + 2 P_{({\rm tr})} +
P_{(3)} \right) = 0  \,. \label{iso5}
\end{equation}
Equations (\ref{iso2}) and (\ref{iso3}) are compatible when
\begin{equation}
\frac{\ddot{a}}{a} \left(6 q_1 {+} 4 q_2 {+} q_3 \right)  {+} 6
q_1 \left(\frac{\dot{a}}{a} \right)^2 {=} 1 - \frac{P_{(3)}-
P_{({\rm tr})}}{B^2(t)} \,. \label{iso6}
\end{equation}
In the minimally coupled case compatibility requires $P_{(3)} =
P_{({\rm tr})} + B^2$. This excludes an isotropic matter
configuration $P_{(3)} = P_{({\rm tr})}$ together with a
non-vanishing magnetic field. For the present non-minimal coupling
the situation is different. Here, the compatibility condition
(\ref{iso6}) takes the form
\begin{equation}
\dot{H}(t) Q_1 + H^2(t) Q_2 = 1 \,, \quad H(t) =
\frac{\dot{a}(t)}{a(t)} \,, \label{compacti}
\end{equation}
where
\begin{equation}
Q_1  = 6q_1 + 4 q_2 + q_3  \,, \quad Q_2  = 12q_1 + 4 q_2 + q_3
\,. \label{compactQ}
\end{equation}
(In the present isotropic case we have $H_a = \frac{\dot{a}}{a} =
H_b = \frac{\dot{b}}{b} = H_c = \frac{\dot{c}}{c} = H$). The
solutions of this compatibility condition can be classified as
follows.

\noindent {\it Case $Q_1 = 0$}

\noindent The Hubble parameter is constant and satisfies the
condition $6 H^2 q_1 = 1$.

\noindent {\it Case $Q_1 \neq 0$}

\noindent For positive $Q_2$ the function $H(t)$ satisfies the
relation
\begin{equation}
\frac{1 - \sqrt{Q_2} H(t)}{1 + \sqrt{Q_2} H(t)} = \frac{1 -
\sqrt{Q_2} H(t_0)}{1 + \sqrt{Q_2} H(t_0)} e^{- \frac{2
\sqrt{Q_2}(t-t_0)}{Q_1}} \,. \label{compacti1 }
\end{equation}
For the subcase $H(t_0)= \frac{1}{\sqrt{Q_2}}$, there exists a
special constant solution
\begin{equation}
H^2(t) = H^2(t_0) = \frac{1}{(12 q_1 + 4q_2 + q_3)} \,.
\label{compacti3}
\end{equation}
For negative $Q_2$ the Hubble function is
\begin{equation}
H(t) \frac{1}{\sqrt{|Q_2|}} \tan{\left\{ \arctan{
\sqrt{|Q_2|}H(t_0)} + \frac{\sqrt{|Q_2|}(t-t_0)}{Q_1} \right\} }
\,. \label{compacti2 }
\end{equation}
For $Q_2 = 0$ the Hubble parameter is linear in time
\begin{equation}
H(t) = H(t_0) - \frac{t-t_0}{6 q_1} \,. \label{compacti4 }
\end{equation}

\noindent Furthermore, for non-minimal coupling the richer
structure of the field equations admits a configuration for which
a non-zero magnetic field does not appear in Einstein's equations.
This happens if all the multipliers of the terms $\kappa B^2(t)$
in (\ref{iso1})-(\ref{iso3}) vanish simultaneously. The
corresponding conditions are
\begin{equation}
H(t) \equiv \frac{\dot{a}}{a} = {\rm const} \equiv H_0 \,, \quad
q_1 = \frac{1}{2 H^2_0} \,, \quad 4q_2 + q_3 = - \frac{5}{H^2_0}
\,. \label{iso7}
\end{equation}
It is interesting to realize that this solution coincides with
(\ref{compacti3}). The third relation admits the particular case
\begin{equation}
q_2
= - \frac{2}{H^2_0} \,, \quad q_3 = \frac{3}{H^2_0} \,,
\label{iso8}
\end{equation}
for which $6q_1 + 3q_2 + q_3 = 0$  and the trace of the
susceptibility tensor vanishes (see, eq.~(\ref{zerotrace})). As a
consequence, there exists a stationary cosmological solution with
$$
P_{(3)}= P_{({\rm tr})} \equiv P \,, \quad  W(t) = const \equiv
W_0 \,, \quad P = {\rm const} = - W_0  \,,
$$
\begin{equation}
3 H^2_0 = \kappa W_0 + \Lambda \,, \quad H(t) = H_0 \,,  \quad
a(t) = a(t_0) e^{H_0(t-t_0)}  \,. \label{iso9}
\end{equation}
The non-vanishing magnetic field is {\it hidden} as far as the
space-time evolution is concerned. The additional coupling terms
give rise to a {\it non-minimal screening} of the magnetic field.

Also the solution (\ref{iso9}) with (\ref{iso7}) and (\ref{iso8})
is characterized by a Hubble expansion that is directly determined
by the non-minimal coupling strength. While this solution  does
not constitute a real inflationary model since there is no exit
from the de Sitter phase, we hope that it may provide the starting
point for a more general approach in which the parameters $q_1$,
$q_2$ and $q_3$ are no longer constants but dynamical degrees of
freedom, e.g., a multiplet of scalar fields. In such a context
more ``realistic" solutions might well occur. The circumstance
that the impact of the non-minimal coupling weakens in the long
time limit becomes obvious if we try to solve, say
eq.~(\ref{iso3}), with a power law ansatz $a \propto t^{\nu}$. All
the terms in the braces on the right hand side, except for the
first one which is independent of the non-minimal coupling, decay
as $t^{-2}$. Consequently, these terms play a role at early times
but they become irrelevant in the long time limit.

\section{Discussion}
\label{Discussion}

Inspired by a well motivated non-minimal coupling between gravity
and electromagnetism we have explicitly demonstrated that the
richer structure of the corresponding theory gives rise to novel
features of the cosmological dynamics. We have obtained a number
of simple exact solutions for Bianchi I models with magnetic
field. For axially symmetric configurations we found inflationary
type solutions with magnetic field which describe an
isotropization process as a result of the non-minimal coupling,
i.e., without a counterpart in the minimally coupled theory.
Furthermore, some solutions of the non-minimal theory establish a
direct relation between a cosmological constant and the coupling
parameters of the non-minimal interaction. Finally, we have shown
that there exists an isotropic de Sitter solution for which the
magnetic field is screened by the non-minimal coupling.

\section*{Acknowledgments}
This work was supported by the Deutsche Forschungsgemeinschaft.

\end{document}